\documentclass[aps,prb,superscriptaddress,twocolumn]{revtex4}
\usepackage{amsmath}
\usepackage{amssymb}
\usepackage{amsthm}
\usepackage{graphicx}
\usepackage{graphics}
\usepackage{subfigure}
\usepackage{bm}
\usepackage{hyperref}
\usepackage{rotating}

\renewcommand{\vec}[1]{\mathbf{#1}}

\newcommand{\abs}[1]{\left\vert#1\right\vert}

\def\beas{\begin{eqnarray*}}
\def\eeas{\end{eqnarray*}}
\def\bea{\begin{eqnarray}}
\def\eea{\end{eqnarray}}
\def\be{\begin{equation}}
\def\ee{\end{equation}}

\newcommand{\bpm}{\begin{pmatrix}}
\newcommand{\epm}{\end{pmatrix}}
\newcommand{\bmm}{\begin{matrix}}
\newcommand{\emm}{\end{matrix}}

\newcommand{\citeasnoun}[1]{Ref.~\onlinecite{#1}}

\begin{document}

\title{Is the electrostatic force between a point charge and
a neutral metallic object always attractive?}

\author{Michael Levin}
\affiliation{Department of Physics, Harvard University, Cambridge MA 02138}

\author{Steven G. Johnson}
\affiliation{Department of Mathematics, Massachusetts Institute of 
Technology, 
Cambridge MA 02139}

\date{\today}

\begin{abstract}
We give an example of a geometry in which the electrostatic force between 
a point charge and a neutral metallic object is repulsive. The example
consists of a point charge centered above a thin metallic hemisphere,
positioned concave up. We show that this geometry has a repulsive regime
using both a simple analytical argument and an exact calculation for
an analogous two-dimensional geometry. Analogues of this 
geometry-induced repulsion can appear in many other contexts, 
including Casimir systems.
\end{abstract}


\maketitle


\section{Introduction}

\begin{figure}[t]
\centerline{
\includegraphics[width=1.0\columnwidth]{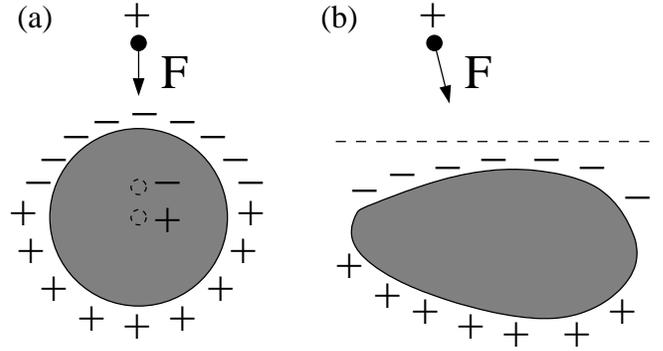}
}
\caption{(a) Electrostatics of a point charge interacting with a 
neutral metallic sphere. Using image charges (dotted circles) it
is easy to see that the force $\vec{F}$ on the point charge is 
attractive. (b) In this paper, we ask whether the force is attractive for 
\emph{any} shape of metallic object. More precisely, if the point charge 
and the object lie on opposite sides of the $z=0$ plane (dotted line), with
the charge in $\{z>0\}$ and the object in $\{z<0\}$, is $F_z$ always 
negative?}
\label{esgen}
\end{figure}

A classic problem in electrostatics is to compute the force between 
a point charge and a perfectly conducting, neutral metallic 
sphere [Fig. \ref{esgen}(a)]. The problem can be easily solved 
using the method of images. One finds that the force on the point 
charge can be computed by summing the forces exerted by
two image charges---one located at the center of the sphere, 
carrying like charge, and one closer to the surface, 
carrying opposite charge. 

A simple corollary of this calculation is that 
the force is always attractive, since the oppositely charged image 
charge is always closer to the point charge than its partner. This 
makes sense intuitively, since one expects that a positively charged 
point charge will induce negative charges on the part of the sphere
that is closest to it, and positive charges on the part that is 
further away. In fact, from this point of view, it is natural to wonder if this 
phenomena is more general and the force is attractive for \emph{any} 
geometry, not just a sphere. 

This question is the main subject of this 
paper. In some sense, one can think of this as an attempt to strengthen Earnshaw's
theorem: recall that Earnshaw's theorem and its generalizations
\cite{Griffiths99,B3953} tell us that a point charge can never be trapped in 
a stable equilibrium via electrostatic interactions with a metallic object.
Here, we ask whether one can go further in the case where the metallic 
object is neutral, and show the force is \emph{always attractive}.

To make the question precise, we need to define what we mean 
by an ``attractive" force. To this end, it is useful to make 
the additional assumption that the charge and metallic object 
lie on opposite sides of a plane, say, the $z=0$ plane, with the 
charge in the upper half space $\{z > 0\}$ and the metal object in 
the lower half space $\{z < 0\}$ [Fig. \ref{esgen}(b)]. Then, by an 
``attractive" force we mean a force $\vec{F}$ on the charge with $F_z < 0$.
 

Given this definition, it is not hard to show that the force is attractive 
in a number of cases. The first case is if the point charge is very close
to the surface of the metal object. In this case, the problem reduces to 
the standard system of a charged particle and an infinite metal plate, which
clearly has an attractive force. The second case is if the point charge is 
very far from the metallic object---say at position $(0,0,z)$ where $z$ is large. 
To see that the force is attractive in this case, recall 
Thomson's theorem \cite{Jackson98}: the induced charges in a metallic 
object always arrange themselves to minimize the total electrostatic 
energy of the system. A corollary of this is that the electrostatic 
energy of a system composed of a metallic object and a charge is always 
lower than the energy of the charge in vacuum. Letting $U(z)$ denote the 
electrostatic energy when the charge is at position $(0,0,z)$, we conclude
that $U(z) \leq U(\infty)$ so that $F_z = -dU/dz$ must be negative (i.e. 
attractive) for large $z$. In addition, one can show that the force is attractive
at any distance, in the case where the metal object is replaced by a 
dielectric material with a dielectric constant $\epsilon = 
1+ \delta$, $0 < \delta \ll 1$. One way to see this is to note 
that, to lowest order in $\delta$, the electrostatic 
interaction between the charge and the object can be decomposed 
into a sum of independent interactions with infinitesimal patches 
of dielectric material. One can then check that each patch gives rise to
an attractive interaction, so that the total interaction is necessarily
attractive. As a final example, one can show that the force is attractive if the 
metallic object is grounded rather than neutral: in that case, a positively
charged point charge will only induce negative charges on the metallic
object, leading to an attractive force.

Given all of this evidence, one might think that the force is always 
attractive. Surprisingly, this is \emph{not} the case. In this paper, we 
give a simple example of a geometry in which a neutral metallic object 
\emph{repels} a point charge. We establish repulsion using both a simple 
analytical argument and an exact calculation for an analogous 
two-dimensional (2d) geometry. In accordance with Earnshaw's theorem and its 
generalizations, \cite{Griffiths99,B3953} our geometry does not yield any 
stable equilibria. However, the fact that one can have repulsion at all is 
surprising, and we show that analogues of this unusual geometric effect 
exist in several other contexts, including Casimir systems.

This paper is organized as follows. In section \ref{repgeom}, 
we describe the counterexample geometry and show that it has a repulsive 
regime. In section \ref{geomor}, we investigate the origin of the 
repulsion and in section \ref{ex2d} we present an exact solution of 
an analogous 2d geometry. Finally, in section 
\ref{gener}, we discuss generalizations and analogues of this unusual
geometric effect.

\section{Example of a repulsive geometry}
\label{repgeom}

\begin{figure}[t]
\centerline{
\includegraphics[width=0.5\columnwidth]{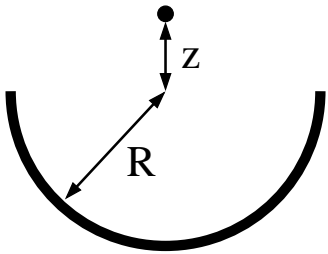}
}
\caption{Example of a geometry in which a neutral metallic 
object repels a point charge: a point charge centered above a thin 
metallic hemisphere (side view). 
}
\label{esexam}
\end{figure}

The geometry that gives a repulsive force consists of an 
thin metallic hemisphere of radius $R$, centered at 
the origin and positioned in the lower half space $\{z<0\}$, 
together with a point charge at some position $(0,0,z)$ on the 
positive $z$ axis (Fig. \ref{esexam}). Note that this system is 
cylindrically symmetric about the $z$ axis, so the force that the 
hemisphere exerts on the charge necessarily points in the $z$ 
direction. When the charge is far from the hemisphere---that is, 
$z \gg R$---the force is necessarily attractive by the general
argument described above. We now show that the force changes sign 
and becomes repulsive when the charge approaches the $z=0$ plane. 

\begin{figure}[t]
\centerline{
\includegraphics[width=1.0\columnwidth]{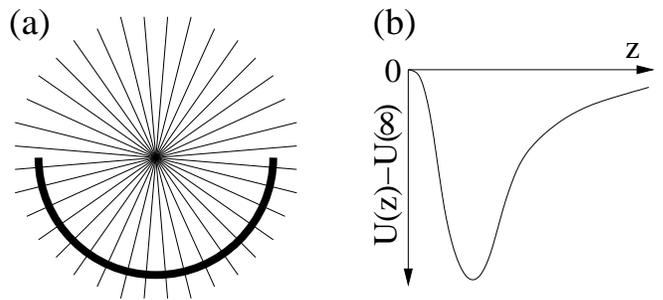}
}
\caption{Argument that a thin metallic hemisphere repels a point charge.
(a) At $z=0$, the vacuum electric field lines of the point charge are 
already perpendicular to the hemisphere (side view), so the electric 
field is unaffected by the presence of the hemisphere. 
(b) Schematic charge-hemisphere interaction energy 
$U(z)-U(\infty)$: zero at $z=0$ and at $z \rightarrow \infty$, and 
attractive for $z \gg R$, so there must be repulsion for small 
positive $z$. 
}
\label{esarg}
\end{figure}

Surprisingly, we can establish the existence of a repulsive regime 
without any calculation at all if we assume an idealized 
geometry where the hemisphere is infinitesimally thin. The basic 
idea is to consider the case where the point charge is at the 
origin, $z=0$. Notice that when the point charge is at this special 
point, the vacuum electric field lines of the point charge are all 
perpendicular to the hemisphere [Fig. \ref{esarg}(a)]. This means 
that the \emph{vacuum} electric field solves the relevant boundary value 
problem. Since the electrostatic energy $U$ of the system is proportional to 
the volume integral of $E^2$, 
\begin{equation}
U = \frac{1}{8\pi}\int E^2 d^3\vec{x}
\label{elener}
\end{equation}
we conclude that the energy of the system is the same as the energy of a 
point charge in vacuum. 
\footnote{Strictly speaking, since the integral in (\ref{elener}) is divergent 
for a geometry with an ideal point charge, we need to be careful to
avoid any references to the absolute energy $U(z)$ (which is infinite)
and instead only consider differences in energies, like 
$U(z)-U(\infty)$. This is implicit in the discussion here.}
In other words, the electrostatic energy at $z=0$ is identical to the 
energy at infinite separation: $U(z=0) = U(z=\infty)$. The 
existence of a repulsive regime now follows since the energy $U$ 
must vary non-monotonically between $z=0$ and $z=\infty$ and hence 
must be repulsive at some intermediate points [Fig. \ref{esarg}(b)].

In fact, we can go a bit further and argue that there is a 
repulsive regime when $z$ is small and positive. Indeed, recall
the inequality $U(z) \leq U(\infty)$ derived in the introduction.
Since $U(0) = U(\infty)$, we have the inequality $U(z) \leq U(0)$ 
implying that $F_z = -dU/dz$ is positive (repulsive) for small 
positive $z$.

As the force is attractive for large $z$ and repulsive for small 
$z$, the simplest consistent scenario is that the electrostatic energy 
$U(z)-U(\infty)$ is zero at $z=0$, decreases to negative values for 
small $z>0$ then increases to zero for large $z$, as depicted in
Fig. \ref{esarg}(b). We confirm this scenario in section 
\ref{ex2d} with an exact solution of an analogous 2d system. 

Note that the point of minimum $U$ is an equilibrium position, stable 
under perturbations in the $z$ direction. By Earnshaw's theorem and
its generalizations \cite{Griffiths99,B3953}, this equilibrium point must be
unstable to lateral (xy) perturbations. More generally, using
the exact 2d solution (\ref{potgen}), one can show that the point charge is 
unstable to (xy) perturbations at \emph{all} points on the $z$ axis.

So far, we have focused on an idealized geometry where the 
hemisphere is infinitesimally thin. Now suppose that the 
hemisphere has a finite thickness $t$. In this case, it is no
longer true that $U(0) = U(\infty)$ and hence the above argument 
cannot be applied directly. However, as long as $t/R$ is small, the 
repulsive regime must persist since the electrostatic energy curve 
[Fig. \ref{esarg}(b)] can only shift by a small amount from the 
$t=0$ case. To make this more quantitative, note that the 
main effect of the finite thickness is to expel the electric field from a 
finite volume $V = t \cdot 2\pi R^2$. As a result, we have (to 
lowest order in $t$)
\begin{equation}
U(0) - U(\infty) = -\frac{1}{8\pi} E^2 V = -\frac{q^2 t}{4R^2}   
\end{equation}
where $q$ is the charge carried by the point charge. Comparing 
this with the minimum value of $U$, which is of order
$U_{min}-U(\infty) \sim -q^2/R$ at $t=0$, we see that $U_{min} < U(0)$ for small 
$t/R$, so the repulsive regime must persist in the presence of small, finite 
thickness. On the other hand, when $t/R$ becomes sufficiently 
large, the repulsive regime disappears completely, as we explain 
in the next section.

\section{Geometric origin of the repulsion}
\label{geomor}

\begin{figure}[t]
\centerline{
\includegraphics[width=1.0\columnwidth]{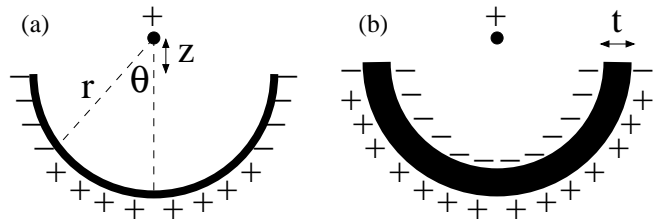}
}
\caption{(a) Schematic charge density $\sigma$ induced by a
point charge on an infinitesimally thin hemisphere (side view). The 
force that an induced charge on the hemisphere exerts on the point 
charge in the $z$ direction is proportional to $\cos{\theta}/r^2$. 
The positive charges have a larger $r$ then the negative charges, 
but also a larger $\cos{\theta}$. The latter effect dominates and 
leads to a repulsive force for small, positive $z$. (b) Induced 
charge density on a hemisphere with finite thickness $t$. The 
displacement between the two surface densities makes an attractive 
contribution to the force and destroys the repulsive regime when 
$t$ is large.}
\label{esqual}
\end{figure}

The general argument described above proves that there must be a 
repulsive regime, but it does not tell us what causes the repulsion. 
To address this question, it is useful to consider the induced charges 
on the hemisphere when the charge is at some point $(0,0,z)$ on the
positive $z$ axis. In general, there will be charges on both sides of the 
hemisphere, but in the limit where the hemisphere is very thin, we 
can make the approximation of combining the charges on the two sides 
into a single surface charge density $\sigma$. Assuming that the 
point charge is positive, we expect this total charge density to be of 
the form shown in Fig.\ref{esqual}(a), with $\sigma$ positive in 
the center of the hemisphere and negative near the boundary. We would 
like to understand the force that these induced charges exert on the 
point charge. Clearly the negative charges are closer to the point
charge than the positive charges, so they exert a stronger force on
it. Naively, one might expect this to lead to a net attractive
force. However, the key point is that the angle between the force
direction and the $z$ axis is smaller for the positive charges
than the negative charges, so even though they are further away, 
they can potentially exert more force in the $z$ direction, 
depending on the position of the point charge.

More precisely, the $z$ component of the force that a charge 
on the hemisphere exerts on the point charge is proportional to 
$\cos{\theta}/r^2$ where $r$ is the distance to the charge, and $\theta$ 
is the angle with respect to the $z$ axis [Fig. \ref{esqual}(a)]. 
The positive charges have a larger $r$, but also a larger 
$\cos{\theta}$ then the negative charges; the competition between 
these two geometrical effects determines the sign of the force. If 
the point charge is very close to the origin, 
$z \ll R$, then the trigonometric factor $\cos{\theta}$ wins out: $r$ is 
virtually the same for the positive and negative charges ($r 
\approx R$), but $\cos{\theta}$ is much larger for the positive 
charges. The result is a repulsive force. On the other hand, if 
the point charge is very far away, $z \gg R$, then $\cos{\theta}$ is 
virtually the same for the positive and negative charges (i.e. 
$\cos{\theta} = 1 + O(R^2/z^2)$) while the $1/r^2$ factor is 
larger for the negative charges (i.e. larger by a 
factor of size $1 + O(R/z)$). The result is an attractive 
force.

This picture also explains why the repulsive regime disappears 
when the thickness $t$ becomes comparable to $R$. Indeed, once
$t/R$ is appreciable, we can no longer make the approximation of 
combining the charges on the two sides of the hemisphere into a 
single charge density. Instead, we need to treat the two 
surface charge densities separately. While the sum of the two 
charge densities has the form shown in Fig. \ref{esqual}(a), we 
expect that the charges on the inner surface are primarily 
negative, while the charges on the outer surface are primarily 
positive, as depicted in Fig. \ref{esqual}(b). The finite 
displacement between the two surfaces makes an attractive 
contribution to the total force, since the negative charges are 
closer to the point charge than the positive charges. This effect 
can overwhelm the $\cos{\theta}$ trigonometric factor when $t/R$ is 
sufficiently large, destroying the repulsive regime completely.

\section{Exact solution in two dimensions}
\label{ex2d}

In this section, we consider a two-dimensional (2d) analogue of the 
repulsive geometry, and solve the associated electrostatics problem 
exactly. (The three-dimensional problem can also be solved exactly, though the
calculation is more involved \cite{Reid10}). The 2d geometry consists of a 
metal semicircle of radius $R$, 
which we denote by $S_R$, together with a point charge. In analogy with 
the three-dimensional (3d) case, we take the semicircle to be centered at the 
origin and positioned in the lower half plane, that is, $S_R = 
\{(x_1,x_2): \abs{x_1}^2 +\abs{x_2}^2 = R^2, x_2 \leq 0\}$, and
the point charge to be at position $\vec{y} = (0,z)$ with $z$ positive. 
 
We now compute the 2d electrostatic interaction between the point 
charge and the metal semicircle assuming that the point charge carries 
charge $q$. Our starting point is the 2d boundary value
problem defined by
\begin{equation}
\nabla^2 \phi_{\vec{y}}(\vec{x}) = 
-2\pi q \cdot \delta(\vec{x}-\vec{y})
\label{2dbvp}
\end{equation}
with the boundary conditions
\begin{align}
\phi_{\vec{y}}(\vec{x}) &= \text{const.} \ \text{for} \ \vec{x}
\in \partial S_R \nonumber \\
\phi_{\vec{y}}(\vec{x})
+q\log{\abs{\vec{x}}} &= 0 \ \text{for} \ \vec{x} \rightarrow
\infty \nonumber \\
\int_{\partial S_R} \vec{n} \cdot \vec{\nabla}
\phi_{\vec{y}}(\vec{x}) d\vec{x} &= 0
\label{2dbc}
\end{align}
Here, the first equation imposes the boundary condition that the
semicircle is an equipotential surface, while the third equation imposes the 
condition that the semicircle is electrically neutral.
The force that the metallic object exerts on the charge is given by
\begin{equation}
\vec{F}(\vec{y}) = -q \vec{\nabla}
\tilde{\phi}_{\vec{y}}(\vec{x})|_{\vec{x} = \vec{y}}                
\end{equation}
where
\begin{equation}
\tilde{\phi}_{\vec{y}}(\vec{x}) =
\phi_{\vec{y}}(\vec{x})+q \log{\abs{\vec{x}-\vec{y}}}                 
\label{2dindpot}
\end{equation}
is the potential created by the induced charges on the metal
object. The electrostatic energy of the system, $U(\vec{y})$, is given by
\begin{eqnarray}
U(\vec{y})-U(\infty) 
&=& -\int_{\infty}^{\vec{y}} \vec{F}(\vec{x}) \cdot d\vec{x} \nonumber \\
&=& \frac{q}{2} \tilde{\phi}_{\vec{x}}(\vec{x}) |_{\infty}^{\vec{y}} 
\nonumber \\
&=& \frac{q}{2} \tilde{\phi}_{\vec{y}}(\vec{y})
\label{2den}
\end{eqnarray}
Here the second equality follows from the fact that 
$\tilde{\phi}_{\vec{x}}(\vec{y}) = \tilde{\phi}_{\vec{y}}(\vec{x})$ so 
that $\frac{q}{2}\vec{\nabla}_x \tilde{\phi}_{\vec{x}}(\vec{x}) 
= -\vec{F}(\vec{x})$.

Our strategy will be to solve the boundary value problem 
(\ref{2dbvp}) using a conformal mapping, obtain 
$\tilde{\phi}_{\vec{y}}$, and then compute the 
energy (\ref{2den}).
To this end, let us view our 2d system as the complex plane 
$\mathbb{C}$, and use complex coordinates $u = x_1+ix_2$, 
$v=y_1+iy_2$ in place of $\vec{x}, \vec{y}$. One can check 
that the analytic function
\begin{equation}
h(u) = \frac{iR+u+i\sqrt{R^2-u^2}}{2}
\end{equation}
defines a conformal map from the region outside the semicircle, 
$\mathbb{C}\setminus S_R$ to the region outside the disk $D$ of 
radius $R/\sqrt{2}$ centered at the origin, $\mathbb{C}\setminus D$. 

The boundary value problem for a metallic disk can be easily solved
using image charges. The potential for this geometry is given by
\begin{equation}
\phi^D_{v}(u) = -q \log{\abs{u - v}} 
+ q \log{\abs{u - \frac{R^2}{2\bar{v}}}} - q \log{\abs{u}}
\end{equation}

\begin{figure}[t]
\centerline{
\includegraphics[width=1.0\columnwidth]{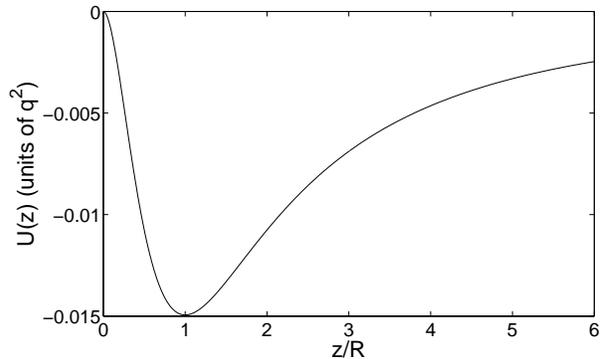}
}
\caption{Exact 2d electrostatic interaction energy $U(z)$ for  
charge-semicircle geometry. In analogy with the 3d case, the metallic 
semicircle is centered at the origin and positioned in the lower half 
plane, while the point charge is at position $\vec{y} = (0,z)$ with $z$ 
positive.}
\label{exact2denergy}
\end{figure}

It follows that the potential for the semicircle geometry is
\begin{eqnarray}
\phi_{v}(u) &=& \phi^D_{h(v)}(h(u)) \nonumber \\
&=& -q \log{\abs{h(u) - h(v)}} + q \log{\abs{h(u) 
- \frac{R^2}{2 h(\bar{v})}}} \nonumber \\
&-& q \log{\abs{h(u)}}
\end{eqnarray}
so that
\begin{equation}
\tilde{\phi}_{v}(v) = -q \log{\abs{\frac{dh}{dv}}} + 
q \log \left(1 - \frac{R^2}{2\abs{h(v)}^2} \right)
\end{equation}  
Substituting in the expression for $h$, we derive
\begin{equation}
\tilde{\phi}_{v}(v) = 
q \log \left( \frac{2- 
\frac{4R^2}{\abs{iR+v+i\sqrt{R^2-v^2}}^2}}{\abs{1-i\frac{v}
{\sqrt{R^2-v^2}}}}\right)
\label{potgen}
\end{equation}
Specializing to the case where $v$ is on the positive imaginary 
axis, $v = iz$, so that $\vec{y} = (0,z)$, and using the convention 
that $U(\infty)=0$, we obtain the electrostatic energy:
\begin{equation}
U(z) = \frac{q^2}{2} 
\log \left( \frac{2- \frac{4R^2}{(R+z+\sqrt{R^2+z^2})^2}}
{1+\frac{z}{\sqrt{R^2+z^2}}}\right)
\end{equation}
A plot of $U(z)$ is shown in Fig. \ref{exact2denergy}. We can see from 
the figure (or from a little algebra) that the force 
$F_z = -\frac{dU}{dz}$ is repulsive for $z < R$ and attractive for 
$z > R$, with $F_z$ vanishing at $z=R$. Using 
(\ref{potgen}), one can check that the equilibrium at $z=R$ is unstable to 
perturbations away from the symmetry axis, as required by Earnshaw's theorem 
and its generalizations. \cite{Griffiths99,B3953} More generally, this instability 
persists for all $z$, not just $z=R$.

As an aside, we note that the vanishing of $F_z$ at $z=R$ can be 
established without any calculation at all: it follows from a simple 
geometric argument similar to the one in section \ref{geomor}. To see this,
consider the $z$ component of the force that an induced charge on the semicircle 
exerts on the point charge. In analogy with Fig. \ref{esqual}(a), this 
quantity is proportional to $\cos{\theta}/r$ where $r$ is the distance to 
the point charge, and $\theta$ is the angle with respect to the $z$ 
axis. For most locations of the point charge, this geometric factor 
varies from place to place on the semicircle, so the force that an induced 
charge exerts on the point charge depends on where it is located. However, 
when the point charge is at exactly at position $(0,R)$, a little 
geometry shows that $\cos{\theta}/r \equiv 1/(2R)$ for \emph{every} point 
on the semicircle. This means that all the induced charges exert 
the same force in the $z$ direction. Since the object is neutral, 
the contributions from the positive and negative induced charges 
cancel exactly and we conclude that $F_z = 0$.

\section{Related phenomena}
\label{gener}

\subsection{A metallic object that repels an electric dipole}
In this section, we give an example of another unusual 
electrostatic geometry: a metallic object that repels an
electric dipole. As in the point charge case, this effect is 
quite counterintuitive. In most cases the interaction between
a dipole and a metallic object is attractive: one needs a special
geometry to get a repulsive force.

\begin{figure}[t]
\centerline{
\includegraphics[width=1.0\columnwidth]{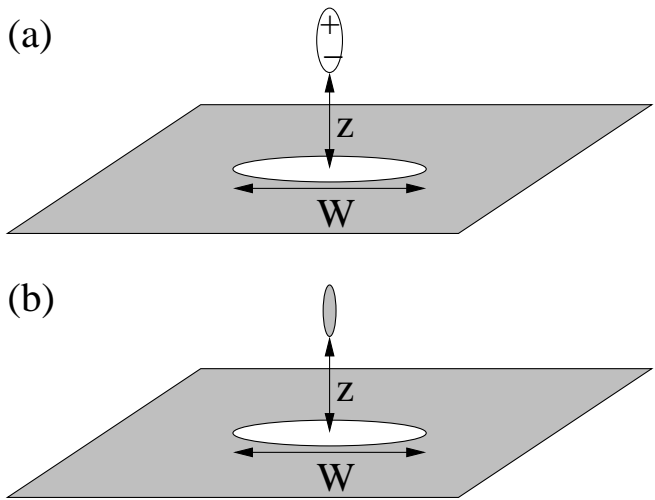}
}
\caption{
(a) Example of a geometry in which a neutral metallic 
object repels an electric dipole: a dipole centered above a thin metallic
plate with a hole. (b) Example of a geometry achieving Casimir repulsion:
an elongated metal particle centered above a thin metal plate with a 
hole.}
\label{esdip}
\end{figure}

The counterexample geometry consists of a metallic plate with a
circular hole of diameter $W$, located in the $z=0$ plane and centered at 
the origin, together with a $z$-directed dipole at position $(0,0,z)$ 
[Fig. \ref{esdip}(a)].

To see that this system has a repulsive regime, we use the same
argument as before: we consider the special case where
the dipole is located at the origin, $z=0$. When the dipole is at
this special point, the vacuum dipole field lines are all
perpendicular to the metal plate. This means that vacuum electric field 
solves the relevant boundary value problem. Since the electric field 
for $z=0$ is identical to the field in vacuum ($z=\infty$), we conclude 
that the energy $U$ is also identical: $U(0) = U(\infty)$. As before, 
this implies that the energy is non-monotonic and hence must be 
repulsive at some intermediate points. Note that the key property of this
geometry is that the metal plate is an equipotential surface for the 
dipole at $z=0$, just as the hemisphere was an equipotential surface for a
point charge at $z=0$.

Again one expects the force to be attractive for large $z$, and 
repulsive for small $z$ so that the energy $U(z)-U(\infty)$ is of the form
shown in Fig. \ref{esarg}(b). One can confirm this picture by exactly
solving a 2d analogue of this geometry (the 3d case can also be
solved exactly, though the calculation is more involved \cite{Reid10}).
The 2d analogue consists of a metal line with a gap of width $W$ located at 
$\{(x_1,x_2) : x_2 = 0, |x_1| \geq 
W/2\}$, together with an electric dipole at $(0,z)$, oriented in the 
$z$-direction. A conformal mapping approach similar to the one in 
section \ref{ex2d} gives
\begin{equation}
U(z) = -p_z^2 \cdot \frac{2 z^2}{(W^2+4z^2)^2}
\end{equation}
(where we are using the convention $U(\infty) =0$. Taking the derivative
with respect to $z$, one finds that the force is attractive for $z > W/2$
and repulsive for $0 < z < W/2$. 

As in the point charge case, one can show that the equilibrium at $z = W/2$
is unstable to perturbations away from the symmetry axis, as required by 
Earnshaw's theorem and its generalizations. \cite{Griffiths99,B3953} More 
generally, one can check that this instability persists for all $z$, not 
just $z = W/2$.

\subsection{A geometry with a repulsive Casimir force}
The Casimir force arises from quantum fluctuations in 
the electric and magnetic polarization of matter. \cite{Landau:stat2} 
It can be regarded as a generalization of the van der Waals force to include 
retardation effects. Most famously, it gives rise to an attractive interaction 
between parallel neutral metallic plates in vacuum. 

A longstanding question is whether the Casimir force between metallic objects
in vacuum is \emph{always} attractive. Using the dipole-metallic object system 
discussed in the previous section, we can show that this is not the case
and construct a simple repulsive geometry for the Casimir force. In the following,
we will describe the geometry and briefly explain why it's repulsive and 
how it's connected to the dipole system. A more detailed discussion can be
found in \citeasnoun{LMR1087}.

The repulsive Casimir geometry consists of a metallic plate with a 
circular hole of diameter $W$, located in the $z=0$ plane and centered at 
the origin, together with an elongated metallic particle at position 
$(0,0,z)$, oriented with the long axis in the $z$ direction 
[Fig. \ref{esdip}(b)]. Our claim is that this geometry has a repulsive 
regime in the limit that the particle is infinitesimally small and highly 
elongated (the limit of an infinitesimal ``metallic needle.") 

To see this, note that the Casimir interaction can be thought of as a 
electromagnetic interaction between zero-point quantum mechanical charge 
fluctuations on the particle and the associated induced charges on the 
plate. As the particle is highly elongated and infinitesimally small, the 
only charge fluctuations are $z$-directed dipole fluctuations; hence the 
problem reduces to understanding the classical electromagnetic interaction
between these $z$-directed dipole fluctuations and the plate with a hole.

The argument now proceeds exactly as in the electrostatic case: we 
consider the special case where the particle is located at the origin, 
$z=0$. When the particle is at this special point, its dipole fluctuations
do not couple to the plate at all, since the vacuum dipole field lines
are already perpendicular to the plate. This is true for not only zero 
frequency dipole fluctuations (as shown in the previous section), but also 
for finite frequency fluctuations. Indeed, the decoupling between the 
dipole fluctuations and the plate is guaranteed by symmetry since the 
metal plate is symmetric with respect to the $z=0$ mirror plane, while the 
dipole fluctuations are antisymmetric. Since the particle and plate do not
couple, it follows that the Casimir energy at $z=0$ is the same as at 
infinite separation, $U(z=0) = U(z=\infty)$, so that the energy must vary 
non-monotonically and hence must be repulsive at some intermediate points.

For $z \gg W$, the hole in the plate can be neglected, and we must have 
the usual attractive interaction. Therefore we expect the 
interaction energy to be of the form shown in Fig. \ref{esarg}(b), with a 
repulsive regime for small $z$, an attractive regime for large $z$, and a 
sign change for at some $z \sim W$. This expectation is confirmed by 
explicit numerical calculation. \cite{LMR1087}

As in the electrostatic examples, the point of minimum $U$ is an unstable
equilibrium as the particle is unstable to perturbations away from
the symmetry axis. Thus, this geometry does not support stable Casimir 
levitation. This is consistent with the instability theorem of 
\citeasnoun{Rahi10:PRL}---an analogue of Earnshaw's theorem for the Casimir force. 
 
\subsection{Current flow analogues}
In this section we construct analogues of these geometric 
effects involving current flow in a resistive sheet. We show that current 
flows can behave in very counterintuitive ways in certain geometries. 
Our starting point is a perfectly homogeneous infinite resistive sheet with
conductivity $\sigma$. Imagine injecting current $I$ into some point 
$\vec{y}$ and collecting it at the infinitely distant boundary. As long 
as the material is homogeneous, then the current will flow out from the 
injection point in a radially symmetric way with the 
current density given by 
\begin{equation}
\vec{j}(\vec{x}) =  \frac{I (\vec{x}-\vec{y})}{2\pi |\vec{x}-\vec{y}|^2}
\end{equation}

\begin{figure}[t]
\centerline{
\includegraphics[width=1.0\columnwidth]{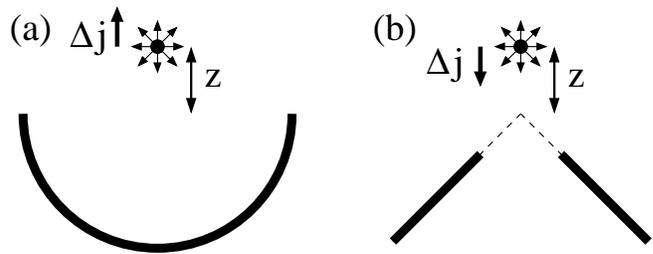}
}
\caption{(a) If current is injected into a homogeneous resistive sheet 
with conductivity $\sigma$, current flows out from the injection point in
a radially symmetric way. Surprisingly, reducing 
the resistivity to $0$ in a thin semi-circular region causes an increase,
$\Delta \vec{j}$, in the current flowing \emph{away} from the semi-circle. 
(b) Increasing the resistivity to $\infty$ along two thin line segments 
intersecting at the origin leads to an increase, $\Delta \vec{j}$, in the
current flowing \emph{towards} the lines.
}
\label{resanalog}
\end{figure}

Consider what happens if one ``shorts out" 
the sheet, reducing the resistivity to $0$ in some 
region $M$. This will break the radial symmetry of the problem and change 
the current flow pattern. Intuitively, one expects that more current will 
flow in the direction of $M$. However this need not be the case: we now
describe a shape $M$ with the property that shorting out
the sheet in $M$ causes current to flow \emph{away} from $M$. 

The counterexample geometry is as follows: one injects current at some
point $\vec{y} = (0,z)$ in the upper half plane, and one shorts out the
sheet along a semicircle centered at the origin and located in 
the lower half plane [Fig. \ref{resanalog}(a)]. When 
$z$ is small, shorting out the sheet along the circle increases the 
current flow in the \emph{positive} $z$ direction in the vicinity of 
$\vec{y}$.

One way to see this is to note that the current 
flow problem can be exactly mapped onto the original electrostatics 
problem. Indeed, the current density $\vec{j}$ obeys the continuum 
analogue of Kirchoff's laws,
\begin{eqnarray}
\vec{\nabla} \cdot \vec{j}(\vec{x}) &=& I \cdot \delta(\vec{x} - \vec{y}) 
\nonumber \\
\vec{\nabla} \times \left(\frac{\vec{j}}{\sigma} \right) &=& 0
\end{eqnarray}
with the boundary conditions
\begin{align}
\vec{j}(\vec{x}) &\perp \partial M \ \text{for} \ \vec{x}
\in \partial M \nonumber \\
\vec{j}(\vec{x}) &= 0 \ \text{for} \ \vec{x} \rightarrow
\infty \nonumber \\
\int_{\partial M} \vec{n} &\cdot \vec{j}(\vec{x}) d\vec{x} = 0
\label{currentbc}
\end{align}
(Here, the first boundary condition comes from the vanishing resistivity
in the region $M$, while the third boundary condition comes from current
conservation). These equations are identical to the equations 
obeyed by the electric field $\vec{E}$ in the point 
charge-metallic object electrostatics problem. But we know that in the 
charge-semicircle electrostatics problem, the metal semicircle generates a
repulsive electric field near the point charge when $z$ 
is small. Translating this into the current flow language, we conclude 
that shorting out the semicircle must increase the current flow in the 
positive $z$ direction, in the vicinity of $\vec{y}$.

It is interesting to consider the opposite question as well: how does 
the current flow change if we cut a hole in the sheet in some region $M$, 
effectively making the resistivity infinite there? Intuitively, one
expects that this will decrease the amount of current flowing towards $M$.
Surprisingly, for some shapes of $M$, this is not the case. 

The counterexample geometry for this problem is to inject current at some
point $\vec{y} = (0,z)$ in the upper half plane and to cut the sheet along
two line segments in the lower half plane, which are symmetric with 
respect to the vertical axis, and which have the property that 
their extensions pass through the origin [Fig. \ref{resanalog}(b)]. When 
$z$ is small, the effect of making these cuts is to increase the current 
flow in the \emph{negative} $z$ direction, at least in the vicinity of 
$\vec{y}$.

To see this, note that in this case, the current density obeys Neumann 
boundary conditions at $\partial M$ instead of Dirichlet boundary 
conditions: 
\begin{align} 
\vec{j}(\vec{x}) &\parallel \partial M \ \text{for} \ \vec{x} \in 
\partial M 
\nonumber \\ 
\vec{j}(\vec{x}) &= 0 \ \text{for} \ \vec{x} \rightarrow \infty 
\label{neucurrentbc} 
\end{align} 
As a result, this current flow problem maps onto a different kind of 
electrostatics problem. Instead of the point charge-metallic object 
problem, the analogue problem in this case involves a point charge and an 
object with a dielectric constant that is much \emph{smaller} than the 
surrounding medium. (Such a geometry is unusual, but could in principle be
realized by immersing a point charge and an object with a small dielectric
constant in a liquid with a large dielectric constant). 

While this electrostatics problem is different from the ones we've 
considered previously, we can analyze it in the same way as before: we 
note that when $z=0$, the vacuum field lines of the point charge 
automatically obey the Neumann boundary conditions (\ref{neucurrentbc}). 
This means that the electric field lines at $z=0$ are the same as in a 
vacuum, so the electrostatic energy $U$ at $z=0$ is the same as at 
infinite separation: $U(z=0) = U(z=\infty)$. Since the force is repulsive 
at large $z$ (this follows from general arguments similar to the 
Dirichlet boundary condition case), we conclude that there is an 
attractive regime at small $z$. Translating this into the current flow 
language, we deduce that cutting the sheet along radial lines 
increases the current flow in the \emph{negative} $z$ direction in the 
vicinity of $\vec{y}$, when $z$ is small.

\section{Conclusion}
In this paper we have shown that, in certain geometries, a neutral metallic
object can repel a point charge. We have also shown that analogues of this 
geometry-induced repulsion can appear in Casimir systems 
and current flow problems. These examples demonstrate that geometry alone can 
reverse the sign of electrostatic and Casimir forces, and lead to 
surprising behavior in many other systems. More generally, we expect that analogues
of this effect can appear in almost any physical system governed by Laplace-like 
equations, from superconductor-magnet systems to (idealized) fluid flow problems.

One direction for future research would be to investigate to what extent these 
counterexamples are special. For example, are all shapes which repel a point charge
similar to the hemisphere geometry discussed here, or are there completely 
different kinds of geometries with this property? More specifically, is it possible
to achieve repulsion with a convex metallic object? One can ask similar 
questions about Casimir repulsion. There are many open questions here---we 
have only just begun to understand these counterintuitive geometric 
effects.

\bibliographystyle{apsrev}
\bibliography{photon}

\begin{thebibliography}{7}
\expandafter\ifx\csname natexlab\endcsname\relax\def\natexlab#1{#1}\fi
\expandafter\ifx\csname bibnamefont\endcsname\relax
  \def\bibnamefont#1{#1}\fi
\expandafter\ifx\csname bibfnamefont\endcsname\relax
  \def\bibfnamefont#1{#1}\fi
\expandafter\ifx\csname citenamefont\endcsname\relax
  \def\citenamefont#1{#1}\fi
\expandafter\ifx\csname url\endcsname\relax
  \def\url#1{\texttt{#1}}\fi
\expandafter\ifx\csname urlprefix\endcsname\relax\def\urlprefix{URL }\fi
\providecommand{\bibinfo}[2]{#2}
\providecommand{\eprint}[2][]{\url{#2}}

\bibitem[{\citenamefont{Braunbek}(2010)}]{B3953}
\bibinfo{author}{\bibfnamefont{W.}~\bibnamefont{Braunbek}},
  \bibinfo{journal}{Z. Phys.} \textbf{\bibinfo{volume}{112}},
  \bibinfo{pages}{753} (\bibinfo{year}{2010}).

\bibitem[{\citenamefont{Griffiths}(1999)}]{Griffiths99}
\bibinfo{author}{\bibfnamefont{D.~J.} \bibnamefont{Griffiths}},
  \emph{\bibinfo{title}{Introduction to Electrodynamics}}
  (\bibinfo{publisher}{Prentice-Hall}, \bibinfo{address}{Upper Saddle River,
  NJ}, \bibinfo{year}{1999}), \bibinfo{edition}{3rd} ed.

\bibitem[{\citenamefont{Jackson}(1998)}]{Jackson98}
\bibinfo{author}{\bibfnamefont{J.~D.} \bibnamefont{Jackson}},
  \emph{\bibinfo{title}{Classical Electrodynamics}}
  (\bibinfo{publisher}{Wiley}, \bibinfo{address}{New York},
  \bibinfo{year}{1998}), \bibinfo{edition}{3rd} ed.

\bibitem[{\citenamefont{Reid}(2010)}]{Reid10}
\bibinfo{author}{\bibfnamefont{M.~T.~H.} \bibnamefont{Reid}},
  \bibinfo{journal}{private communication}  (\bibinfo{year}{2010}).

\bibitem[{\citenamefont{Landau et~al.}(1960)\citenamefont{Landau, Lifshitz, and
  Pitaevski{\u{\i}}}}]{Landau:stat2}
\bibinfo{author}{\bibfnamefont{L.~D.} \bibnamefont{Landau}},
  \bibinfo{author}{\bibfnamefont{E.~M.} \bibnamefont{Lifshitz}},
  \bibnamefont{and} \bibinfo{author}{\bibfnamefont{L.~P.}
  \bibnamefont{Pitaevski{\u{\i}}}}, \emph{\bibinfo{title}{Statistical Physics
  Part 2}}, vol.~\bibinfo{volume}{9} (\bibinfo{publisher}{Pergamon Press},
  \bibinfo{address}{Oxford}, \bibinfo{year}{1960}).

\bibitem[{\citenamefont{Levin et~al.}(2010)\citenamefont{Levin, McCauley,
  Rodriguez, Reid, and Johnson}}]{LMR1087}
\bibinfo{author}{\bibfnamefont{M.}~\bibnamefont{Levin}},
  \bibinfo{author}{\bibfnamefont{A.~P.} \bibnamefont{McCauley}},
  \bibinfo{author}{\bibfnamefont{A.~W.} \bibnamefont{Rodriguez}},
  \bibinfo{author}{\bibfnamefont{M.~H.} \bibnamefont{Reid}}, \bibnamefont{and}
  \bibinfo{author}{\bibfnamefont{S.~G.} \bibnamefont{Johnson}},
  \bibinfo{journal}{arXiv:1003.3487}  (\bibinfo{year}{2010}).

\bibitem[{\citenamefont{Rahi et~al.}(2009)\citenamefont{Rahi, Kardar, and
  Emig}}]{Rahi10:PRL}
\bibinfo{author}{\bibfnamefont{S.~J.} \bibnamefont{Rahi}},
  \bibinfo{author}{\bibfnamefont{M.}~\bibnamefont{Kardar}}, \bibnamefont{and}
  \bibinfo{author}{\bibfnamefont{T.}~\bibnamefont{Emig}},
  \bibinfo{journal}{arXiv:quant-ph/0911.5364v1}  (\bibinfo{year}{2009}).

\end{thebibliography}
\end{document}